\documentstyle[aps,pra,amssymb,psfig]{revtex}

\author{Pawe\l{} Masiak\cite{e-mail}
\footnote[2]{On leave from Centrum Fizyki Teoretycznej Polskiej Akademii Nauk 
and College of Science,\\ Al. Lotnik\'{o}w 32/46, 02-668 Warszawa, Poland}
and Peter L. Knight}
\address{Optics Section, The Blackett Laboratory,
Imperial College, London SW7 2BZ, England}

\title{Copying of entangled states and the degradation of correlations}

\tightenlines

\begin{document}
\maketitle

\begin{abstract}
We investigate the degree to which entanglement survives when a correlated pair
of two--state systems are copied using either local or non--local processes.
We show how the copying process degrades the entanglement, due to a residual 
correlation between the copied output and the copying machine (itself made of
qubits).
\end{abstract}


\section{Introduction}
In classical information theory one can make (in principle) arbitrarily many
perfect copies of any classical input. In quantum information theory, the 
situation is quite different. The no--cloning theorem \cite{no-cloning}
prohibits exact copying of arbitrary superposition states. Nevertheless,
Bu\v{z}ek and Hillery and other authors \cite{copying_1,copying_2,copying_3,copying_4} 
have shown that imperfect copies can be made by a Universal Quantum Cloning Machine (UQCM), 
the outputs of which are identical. 
The price which must be paid is that there is a difference between the original
input and the copies, because of a residual entanglement between machine and 
copies. However not only similarity is lost during the cloning
process. Perhaps even more important than  the no--cloning feature of quantum
mechanics is entanglement, that is a composite microsystem formed from some 
subsystems can be generated in a state $|\phi \rangle$, in which correlations 
between these subsystems are much stronger than any classical correlations. 
This quantum entanglement feature is crucial in many applications, such as
some forms of quantum cryptography \cite{q-crypt1,q-crypt2}, relying on correlated
quantum communication \cite{q-commun} and in quantum computation theory \cite{q-comput}. 
In all these situations success depends upon the strength of the non--local correlations. 
If the cloning process applied to correlated subsystems is to be anything more than a basic 
curiosity, and is to find a practical application in the field of quantum information theory, 
it should be possible to obtain not only maximally accurate copies of the original state, but 
also copies which preserve a degree of non--local correlation characteristic of the copied 
state. We will show in this paper that entanglement is rapidly destroyed by the copying 
process.

\section{Copying}

We consider the simplest generic case of a non--local system, consisting of two qubits 
formed in a pure entangled state. We can choose to describe this in terms of the Bell 
state basis:
\begin{eqnarray}
|\Psi ^{\pm }(\alpha )\rangle &=&\alpha |01\rangle \pm \beta |10\rangle,
\nonumber \\
|\Phi ^{\pm }(\alpha )\rangle &=&\alpha |00\rangle \pm \beta |11\rangle,
\nonumber
\end{eqnarray}
where $\alpha$ determines the amount of entanglement in the state and
$\beta =\sqrt{1-\alpha ^{2}}$. (We assume, for simplicity, that both
$\alpha$ and $\beta$ are real.) 
There are two possibilities: the pair of qubits can be cloned locally or non--locally. 
Now we concentrate on the first case and restrict our considerations to the state 
$|\Psi^{-}(\alpha )\rangle$ as the results for the remaining states are entirely the same.
We will explain below why they are the same for different Bell states. 
\subsection{Local cloning}

A scheme which will achieve local cloning is described by the following\cite{local_cloning}:
two distant parties share an entangled two--qubit state 
$|\Psi ^{-}(\alpha )\rangle$. Each of them perform some local transformations 
on the own qubit using distant quantum copying machines. We assume that the two 
additional qubits, employed in the copying  process, are initially uncorrelated. 
The two copiers make copies of the qubits separately. The result of the cloning 
process is an output which is no longer a pure state, but is described by the 
density matrix:
\begin{equation}
  \hat{\rho}(\alpha)
    =\frac{24\alpha ^{2}+1}{36}|01\rangle \langle 01|
    +\frac{24\beta ^{2}+1}{36}|10\rangle \langle 10|
   \label{local cloning}
\end{equation}
$$
   -\frac{4 \alpha \beta }{36}(|00\rangle \langle 00|+|11\rangle \langle 11|)
    +\frac{5}{36}(|01\rangle \langle 10|+|10\rangle \langle 01|).
$$ 
It is quite easy to determine whether a given pure state is local or non--local. The situation 
is much more complicated for the case when a system is prepared in a mixed state. There is no
general method known to us which determines the degree of non--locality of mixed states.
However in the case of $2 \times 2$ systems, there a is simple criterion which allows us to 
distinguish density matrices describing entangled mixed states \cite{transp1,transp2}.
The Peres-Horodecki separability criterion tells us that if the partially transposed density 
matrix of the composite system, defined by matrix elements generated in the following way 
$\rho^{T_2}_{an,bm}=\rho_{am,bn}$, does not have non--negative eigenvalues, then the state is separable.
For a system formed from two two--level subsystems opposite is also true.
Using the partial transposition criterion we determine for what values of the superposition parameter 
$\alpha$ the final state of the local process is inseparable. The state $\hat{\rho}(\alpha)$ is entangled 
when the following condition is fulfilled: $\frac{1}{2}-\frac{\sqrt{39}}{16}\leq \alpha ^{2}
\leq \frac{1}{2}+\frac{\sqrt{39}}{16}$.

\subsection{Non--local cloning}

The state $|\Psi^{-}(\alpha )\rangle $ can also be cloned non--locally
\cite{register_copying_1,register_copying_2}. 
In this case the entangled state of the two--qubits is treated as a
state in a larger Hilbert space and cloned as a whole. The final state of each
pair of the two--qubit copies at the output of the cloning machine is
given by the density operator: 
\begin{equation}
  \hat{\rho}=\frac{6\alpha^2+1}{10}|01\rangle \langle 01|
    +\frac{6\beta^2+1}{10}|10\rangle \langle 10|
  \label{non-local cloning} 
\end{equation}
$$
    -\frac{3\alpha \beta }{5}(|01\rangle \langle 10| +|10\rangle \langle 01|)
    +\frac{1}{10}(|00\rangle \langle 00|+|11\rangle \langle 11|).
$$
Now, we find the inseparability condition derived from the partial transposition
criterion \cite{transp1,transp2} is the following: $\frac{1}{2}-\frac{\sqrt{2}}{3}\leq \alpha ^{2}
\leq \frac{1}{2}+\frac{\sqrt{2}}{3}$.

\section{Bell inequality}
\label{section_3}
The non--locality of a quantum state can manifest itself in many different ways. 
The best known, and one which has been tested in many experiments is the manifestation
of strong nonclassical correlations of quantum states, in violation of Bell's 
inequality \cite{aspect1,aspect2}. The Bell--CHSH inequality \cite{Bell-CHSH_inequality}
under consideration has the form:
\begin{equation}
{\cal B}=|E(\vec{a},\vec{b})-E(\vec{a^{\prime }},\vec{b})
+E(\vec{a},\vec{b^{\prime}})+E(\vec{a^{\prime }},\vec{b^{\prime }})|\leq 2,
\label{Bell_inequality}
\end{equation}
where
\begin{equation}
E(\vec{a},\vec{b})=\langle \vec{a}\:\vec{\sigma} ^{(1)}
\cdot\vec{b}\:\vec{\sigma}^{(2)}\rangle
=\sum_{i,j=1}^{3} a_{i}b_{j} Tr[\hat{\rho}\:\hat{\sigma}_{i}^{(1)}
\otimes \hat{\sigma}_{j}^{(2)}]
\end{equation}
is the two--qubit correlation function and $\hat{\sigma}_{i}$ are 
Pauli spin-$\frac{1}{2}$ operators. The quantity ${\cal {B}}$ depend
strongly upon vectors $\vec{a}$, $\vec{b}$, $\vec{a^{\prime }}$,
$\vec{b^{\prime }}$. It is easy to check that for the singlet state 
$|\Psi^{-}\rangle = |\Psi ^{-} \left(\frac{1}{\sqrt{2}}\right)\rangle$ 
optimal configuration, maximising ${\cal {B}}$ is
achieved by coplanar vectors $\vec{a}$, $\vec{b}$, $\vec{a^{\prime }}$,
$\vec{b^{\prime }}$, where the angles between two consecutive vectors are
the same and equal to $\pi/4$ (Fig. 1). 
\noindent
Unfortunately this configuration is optimal only for the choice of 
an exact singlet initial state and actually does not fit our needs here. Instead of  
looking for other optimal sets of vectors, which are necessary to calculate the quantity
${\cal {B}}$, we employ the formula for the maximal value of ${\cal {B}}$
obtained recently by Horodecki {\em et.al.} \cite{bell_max_1,bell_max_2}. 
The ${\cal {B}}_{max}$ can be calculated directly using the result 
${\cal {B}}_{max}=2\sqrt{M(\hat{\rho})}$, where $M(\hat{\rho})=max_{i<j}(u_{i}+u_{j})$, 
and $u_{i=1,2,3}$ are eigenvalues of the matrix 
$U(\hat{\rho})=T(\hat{\rho})^{\dagger }T(\hat{\rho})$, and $T_{i,j}(\hat{\rho})
=Tr[\hat{\rho}\hat{\sigma}_{i}^{(1)} \otimes \hat{\sigma}_{j}^{(2)}]$. Using this
expression we investigate properties of copies produced from the pure
entangled state $|\Psi ^{-}(\alpha)\rangle$. These same calculations can be
repeated for the other Bell states, but we have found that the results are
identical to the above and independent of which state was chosen. This results from
the fact that ${\cal {B}}_{max}$ depends upon the state just by the maximum $M(\hat{\rho})$
of sums of pairs of the eigenvalues of the matrix $U(\hat{\rho})$. Clones obtained from
different Bell states $|\Psi^{\pm}(\alpha)\rangle$ or $|\Phi^{\pm}(\alpha)\rangle$ have 
the same eigenvalues $u_i$, so $M(\hat{\rho})$ is the same for all states from Bell basis. 
In Fig. \ref{bell_pure} we compare the values of ${\cal B}$ calculated employing 
vectors $\vec{a}$, $\vec{b}$, $\vec{a^{\prime }}$, $\vec{b^{\prime }}$ taken in
the configuration which is shown in Fig. 1 and values of the ${\cal B}_{max}$, 
both calculated for the Bell state $|\Psi^{-}(\alpha)\rangle$. 
In Fig. \ref{bell_local} and Fig. \ref{bell_non-local} 
we show ${\cal B}$ and ${\cal B}_{max}$ calculated for
states $\hat{\rho}[\Psi(\alpha)]$ of a pair of qubits obtained in the cloning
process, first using local copying and secondly using non--local copying. It turns out
that in both cases the pairs of qubit copies are in states which do not violate
Bell's inequality, even if the input state was prepared in a singlet state with maximal 
entanglement.

\section{Entanglement of formation}

In the previous section we have shown that during the cloning process entangled
states lose so much of their entanglement that they are not able to violate
Bell's inequality in a state correlation experiment. However this does not
mean that such states do not contain any quantum correlations. It is
known that the violation of Bell's inequality is not a wholly satisfactory
measure of quantum non--locality\cite{ent_0}: this is just not as sensitive as other 
measures of entanglement \cite{ent_1,ent_2,ent_3,ent_4,ent_5}. 
There are other quantities which are have been 
developed as measures of quantum correlations and entanglement.
For the case of a pair of correlated qubits, one such measure is the entanglement 
of formation \cite{ent_2,ent_for_1}. This quantity is much
more sensitive to the degree of correlation than Bell's inequality and more importantly,
a finite, compact analytical formula for it is known. The entanglement of formation
$E(\hat{\rho})$ is defined in the following way \cite{ent_2,ent_for_1}:
\begin{equation}
  E(\hat{\rho})={\cal E}(C(\hat{\rho})),
\end{equation}
where
\begin{equation}
  C(\hat{\rho})=\mbox{max}\,\{0,\lambda _{1}-\lambda _{2}-\lambda _{3}-\lambda_{4}\},
\end{equation}
\begin{equation}
  {\cal E}(y)=h\left( \frac{1+\sqrt{1-y^{2}}}{2}\right) ,
\end{equation}
and $h(x)=-x\log _{2}(x)-(1-x)\log _{2}(x)$. The $\lambda _{i}$'s in these
expressions are the square roots of eigenvalues, in decreasing order, of a
non-Hermitian matrix $\hat{\rho} \hat{\tilde{\rho}}$, and $\hat{\tilde{\rho}} 
=(\hat{\sigma}_{y}\otimes \hat{\sigma}_{y})\hat{\rho}^{*}(\hat{\sigma} _{y}
\otimes \hat{\sigma}_{y}).$ 
In Fig. 5 we show the entanglement of formation calculated for the pure state
$|\Psi^{-}(\alpha)\rangle$ and the two--qubit states obtained as the result of 
both local and non--local copying of the state $|\Psi^{-}(\alpha)\rangle$. 
We see that non--local cloning is much more efficient than the local process, 
in that the values of ${\cal E}$ are much larger in the former case. 
The entanglement of formation is in this case non--zero for states 
$|\Psi^{-}(\alpha)\rangle$, characterised by values of $\alpha$ parameter, 
which belong to a larger subinterval of the $[0,1]$ interval.

\subsection{Repetition of non--local cloning}

One sees from the above that either local or non--local copying are rather
inefficient processes from the point of view of preserving entanglement.
Even in the case when the maximally inseparable state
is the input state of the cloning process, only a small amount of entanglement
survives the cloning. It is an interesting question to ask what will happen 
when the output state of the copier is used as an input state in the next step 
of a sequence of cloning processes. In particular, we are interested to discover
how fast the entanglement decreases when the copying is iterated. We restrict 
ourself in these considerations to the case of non--local cloning, because this 
scheme is much more effective as we saw above, and results obtained in this 
case can be treated as an upper bound for all other schemes.
\newline
The final state of the copying process is a mixed state, described by the density 
matrix, eq.(\ref{non-local cloning}). 
This density matrix cannot be used directly as input data in computations, because
the non--local copying scheme works straitforwardly only when an input state is initially 
in a pure, potentially entangled, state. The density matrix should be first converted to a 
form which allows us to perform the second cloning. It turns out that a simple diagonalization
of the density matrix is sufficient. In this new, diagonal base, the density matrix is given 
by the mixture of projection operators $\hat{\rho}=\sum_{i=1}^{4}a_{i}|\phi _{i}\rangle
\langle \phi _{i}|$. The weights $a_{i}$ in the decomposition are the
eigenvalues of the density matrix $\hat{\rho}$ and the vectors
$|\phi_{i}\rangle $ are the normalised eigenvectors of $\hat{\rho}$.
Each vector $|\phi _{i}\rangle $ can be cloned separately. The mixture of the resultant
density matrices taken with the weights $a_{i}$ is the result of the second
cloning process. We repeat this procedure and investigate changes in the
entanglement of formation at each stage of this procedure.
In Fig. 6 we show the entanglement of formation of the clones of the
states $|\Psi^{-}(\alpha)\rangle$ for the case of the first two stages of cloning.
Results for the maximally entangled state $|\Psi^{-}\rangle$ are shown
in Table 1.
\begin{center}
\vspace{0.3cm}
\begin{tabular}{|c|c|c|c|c|}
\hline
N$^{\underline{o}}$ & 0 & 1 & 2 & 3 \\
\hline
\hline
\hspace{0.75cm}${\cal E}$\hspace{0.75cm} & \hspace{0.75cm}1\hspace{0.75cm} & 0.250225 & 0.005094 & \hspace{0.75cm}0\hspace{0.75cm} \\
\hline
\end{tabular}
\vspace{0.5cm}

\noindent TABLE 1. Entanglement of formation of clones of the singlet state 
$|\Psi^{-}\rangle$ \\
as a function of the number of cloning steps.
\end{center}

\noindent One sees the entanglement decreases extremely rapidly, and after just three iterations the copy
of any inseparable state the resultant entanglement of formation goes to zero.

\section{Conclusions}
We have shown that the UQCM of Bu\v{z}ek and Hillery \cite{copying_1} can generate copies of entangled
pairs of qubits, but that the degree of entanglement of the resultant copies is substantially
reduced. The amount of entanglement decreases very rapidly with the number of times the copier is used.
States obtained as a result of either local
or non--local copying do not violate already Bell's inequality. However the entanglement of formation 
of such states remains still positive. It is still positive after the second step of the copying process,
when output states obtained  in the previous step become input states of the next step. But after
the third step the entanglement of formation is 
equal to zero. It means that even qubits,
which are copies of copies of copies of the singlet state are already in local state. They do not have any
nonclassical correlations and are useless as a resource in quantum computation or quantum telecommunication. 
This reduction is of course due to a residual entanglement between the copies and the 
quantum copying machine. This is important if the copying process is used to replicate (albeit 
approximately) copies of a quantum register, for example, in quantum computing \cite{register_copying_1} 
or of a Bell--correlated quantum cryptography protocol \cite{q-commun}.

\section{Acknowledgements}
This work was supported in part by the UK Engineering and Physical Sciences Research Council,
the Royal Society and the European Union. We would like to acknowledge fruitful 
discussions with Daniel Jonathan and Martin Plenio. We also thank Vladimir Bu\v{z}ek
for many useful suggestions and comments on this paper.

\begin{figure}[hbt]
  \centerline{\psfig{figure=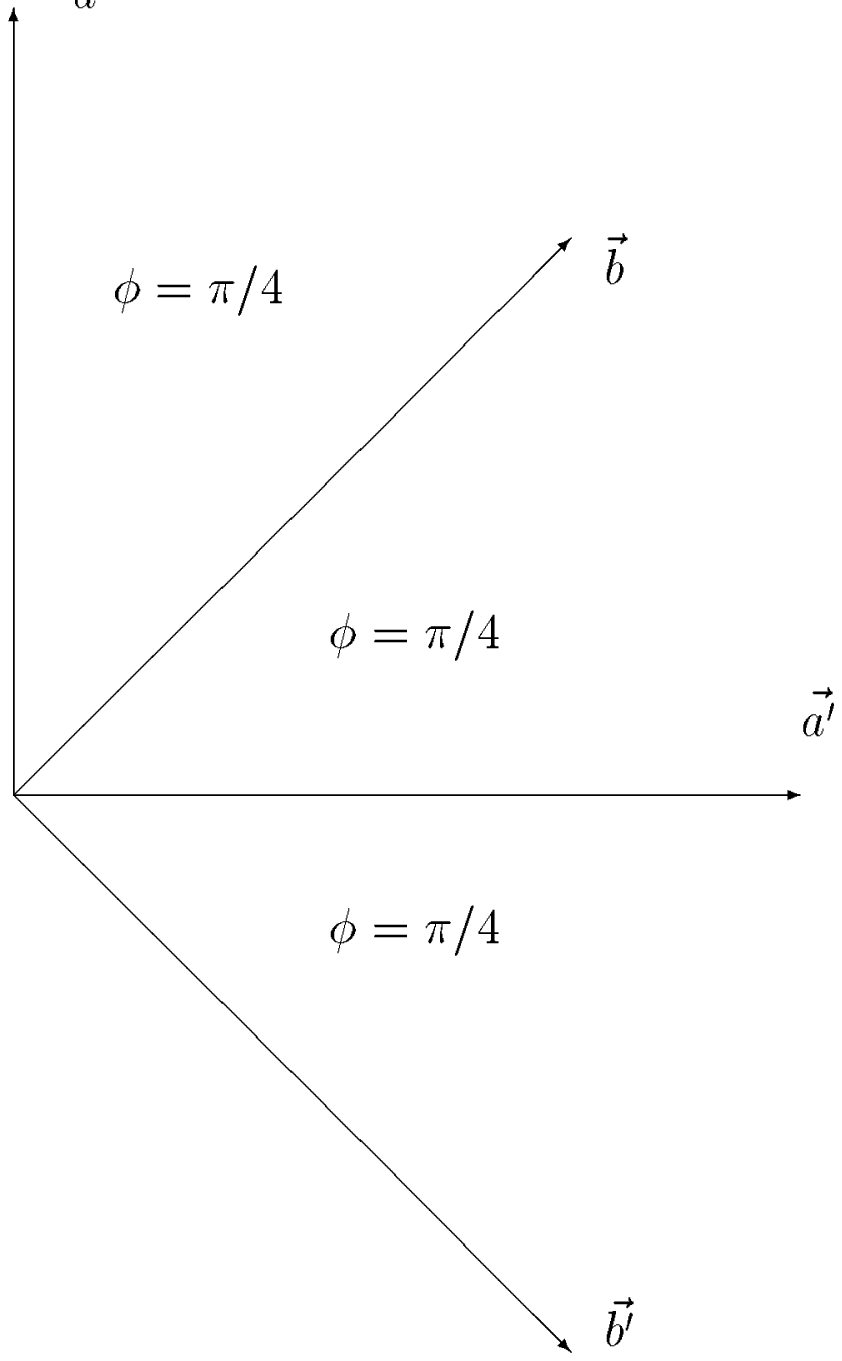,height=20cm}}
  \caption{Optimal configuration of the vectors for the $|\Psi^{-}\rangle$ Bell state.
    \label{vectors}}
\end{figure}
%
\begin{figure}[hbt]
  \centerline{\psfig{figure=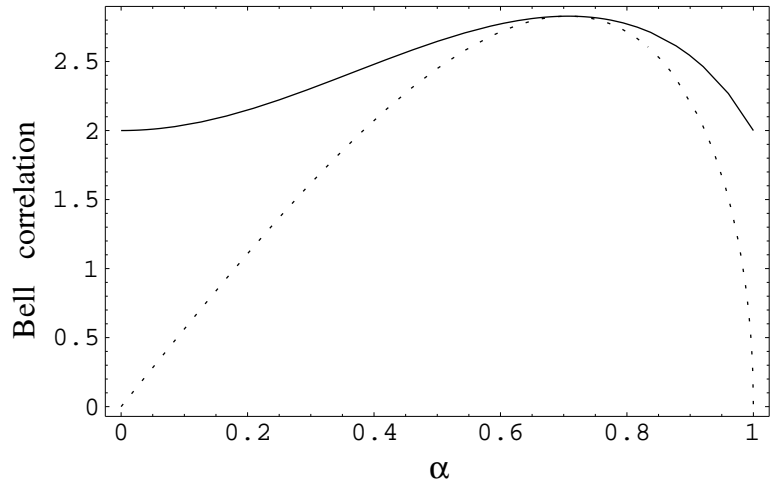}}
  \caption{Bell correlation ${\cal B}$ (defined in eq. \ref{Bell_inequality}) --- dashed curve 
and maximal Bell correlation ${\cal B}_{max}$ (defined in the section \ref{section_3})
--- solid curve, as functions of the superposition amplitude $\alpha$ of states in the Bell
basis for the pure state $|\Psi^{-}(\alpha)\rangle$.
    \label{bell_pure}}
\end{figure}
%
\begin{figure}[hbt]
  \centerline{\hbox{\psfig{figure=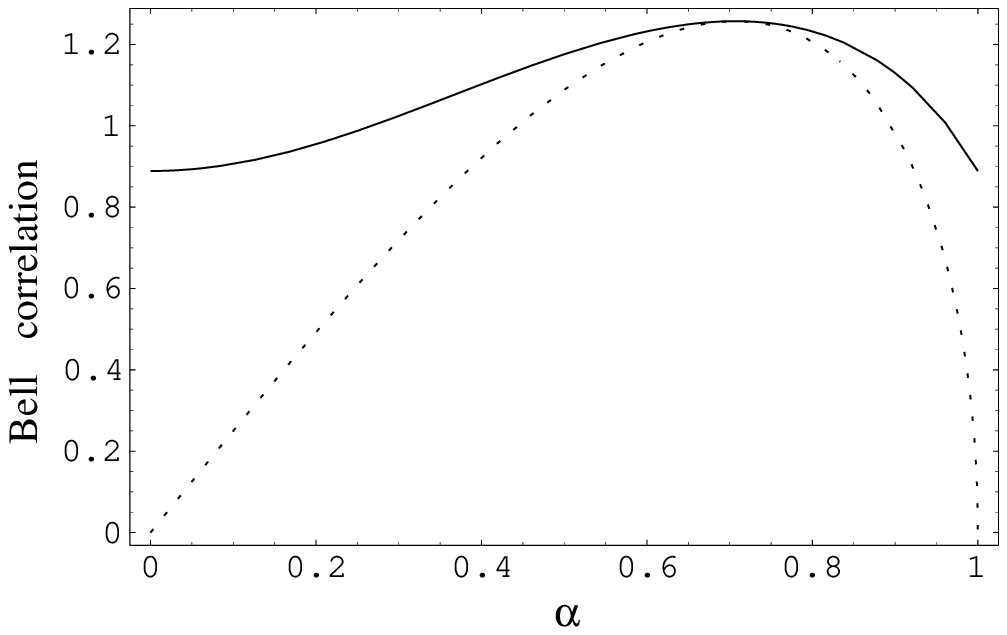}}}
  \caption{Bell correlation ${\cal B}$ (defined in eq. \ref{Bell_inequality}) --- dashed curve 
and maximal Bell correlation ${\cal B}_{max}$ (defined in the section \ref{section_3})
--- solid curve, as functions of the superposition amplitude $\alpha$ of states in the Bell
basis. Local cloning has been employed to obtain the copies.
    \label{bell_local}}
\end{figure}
\newpage
\begin{figure}[hbt]
  \centerline{\psfig{figure=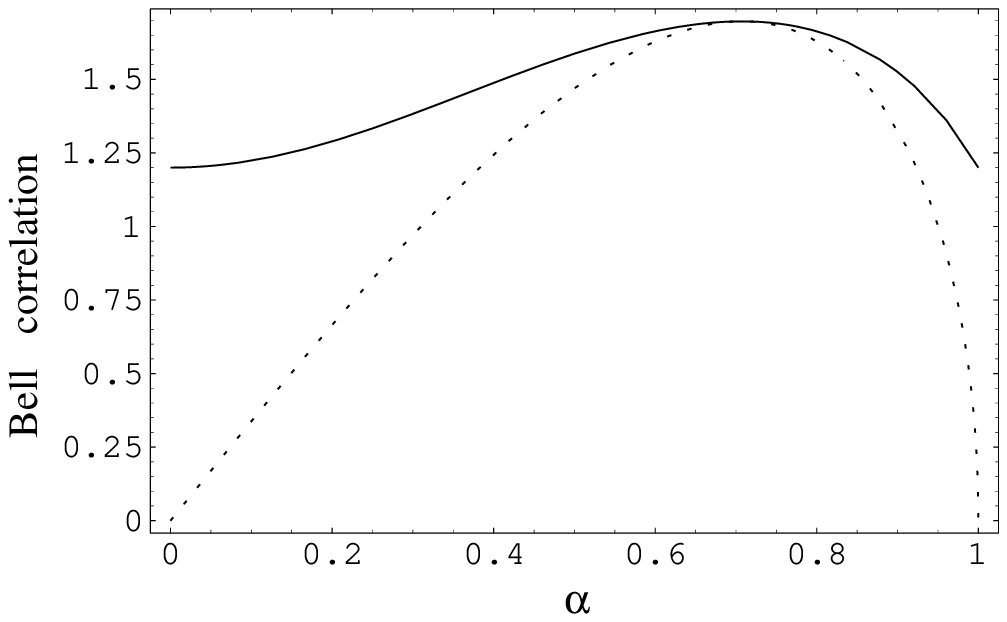}}
  \caption{Bell correlation ${\cal B}$ (defined in eq. \ref{Bell_inequality}) --- dashed curve 
and maximal Bell correlation ${\cal B}_{max}$ (defined in the section \ref{section_3})
--- solid curve, as functions of the superposition amplitude $\alpha$ of states in the Bell
basis. Non--local cloning has been employed to obtain the copies.
    \label{bell_non-local}}
\end{figure}
%
\begin{figure}[hbt]
  \centerline{\psfig{figure=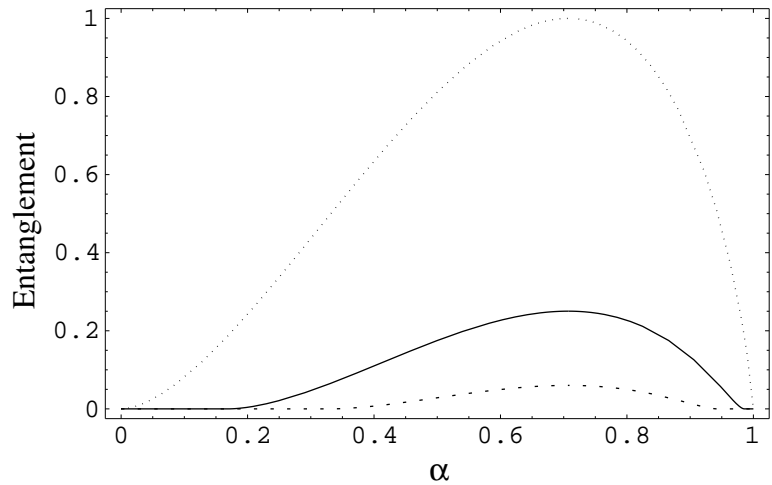}}
  \caption{Entanglement of formation of the pure state $|\Psi^{-}(\alpha)\rangle$ (dotted curve)
and entanglement of formation remaining after the first step of local (dashed curve) and non-local 
(solid curve) cloning.
    \label{ent1}}
\end{figure}
%
\begin{figure}[hbt]
  \centerline{\psfig{figure=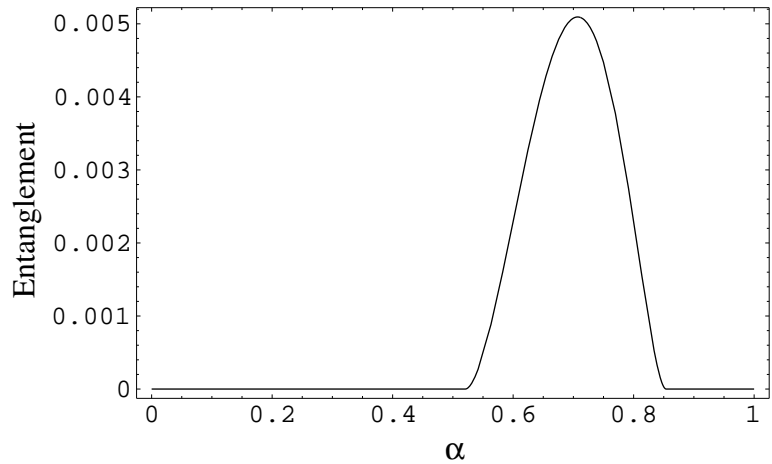}}
  \caption{Entanglement of formation remaining after the second step of non--local cloning.
The case for local cloning is not shown as it is essentially zero.
     \label{ent2step}}
\end{figure}
%

\end{document}